# Comparison of Classification Algorithms for COVID19 Detection using Cough Acoustic Signals

Y.E. ERDOĞAN [1, 2] and A. NARİN [2]

[1] Electronic Automation Department, Eregli Iron and Steel Co., Zonguldak, Turkey,
yeerdogan@erdemir.com.tr
[2] Electrical and Electronics Engineering Department, Zonguldak Bülent Ecevit University, 67100, Zonguldak, Turkey,
alinarin45@gmail.com

*Abstract* - **The epidemic disease, called the new coronavirus (COVID19), firstly occurred in Wuhan, China in December 2019. COVID19 was announced as an epidemic by World Health Organization soon after. Some of the symptoms of this disease are fever, cough, shortness of breath and difficulty in breathing. In more severe cases, death may occur as a result of infection. The most significant question in fighting the pandemic and controlling the epidemic is the early diagnosis of COVID19(+) patients and the follow-up of these patients. Therefore, various diagnostic mechanisms are used. Additionally to the RT-PCR test, medical imaging methods have been utilized, especially in the detection of COVID19(+) patients. In this study, an alternative approach was proposed by using cough data, which is one of the most prominent symptoms of COVID19(+) patients. The cough acoustic public dataset on the Virufy website was used. The entire data was normalized using z-normalization technique. The performance of the features obtained via the 5-layer empirical mode decomposition method and the performances of different classifiers has been compared. As the classifier algorithm, 5 different algorithms were used. The highest accuracy and F1-score performances were obtained by using Ensemble-Bagged-Trees algorithm as 90.6% and 90.5%, respectively. On the other hand, other classification algorithms used in the study are Support Vector Machines, Logistic Regression, Linear Discriminant Analysis and k-Nearest Neigbors, respectively. According to the results obtained, choosing the right classifier algorithm provides high results. Thus, it is clear that using cough acoustic data, those with COVID19(+) can be detected easily and effectively.**

## I. Introduction

By October, 26 2021, there have been 243.572.402 confirmed cases of COVID19, including 4.948.434 deaths, reported to World Health Organization (WHO) and it is also disturbing life in most of the countries and territories all over the world [1]. This virus gives rise to serious respiratory infections with really high deadness and creates serious threats to people. Some symptoms of the COVID19 pandemic are serious fever, dry cough, and difficulty in breathing [2]. In addition to these signs, a range of tests have been done to find out COVID19(+) sicks. One of these tests is Reverse Transcription Polymerase Chain Reaction (RT-PCR) test known as the gold standard, and it is applied with advice of WHO [3]. But it is boring, costly and outcomes are achieved so late [4]. The other methods are medical imaging techniques which utilized to determine COVID19(+) sicks in addition to this test [5]. Researches based on cough acoustic signal analysis has lately taken its place in the literature.

Sharma et al. worked on a dataset which is named "Coswara" including signal data for cough, breath, and voice. They recommended a research with 28 spectral measurements and random forest classifiers. The general accuracy rate got was 67.7% [6]. In another research, Imran et al. purposed to detect Coronavirus sicks by seeking at these data with research they named AI4COVID19. They utilized mel-spectrogram images for convolutional neural network models. They engaged Mel Frequency Cepstral Coefficient (MFCC) and Principal Component Analysis (PCA) based on feature extraction and SVM classification algorithm for the traditional machine learning approach. As a conclusion, they had 95.6% accuracy and 95.6% F1-score [7]. Erdoğan and Narin utilized cough acoustic signals in their research. They utilized z-normalization as preprocessing method. In the research, they utilized intrinsic mode function (IMF) and discrete wavelet transform (DWT) based feature extraction via traditional machine learning approaches and they utilized SVM as a classification method. The overall accuracy for traditional machine learning approaches was 98.4% [8]. They also added deep features to this study. They used deep features on the ResNet50 deep learning model and obtained 97.8% overall accuracy.

In the studies mentioned above, researches using different voice data, especially cough data, were included. In this study, time domain and nonlinear features were obtained by using the 5-layer empirical mode decomposition (EMD) technique. The performances of five different classification algorithms were compared using the obtained features. The 5-layer EMD method used in cough acoustic signals of COVID19(+) and COVID19 (-) persons is shown in Figure 1.

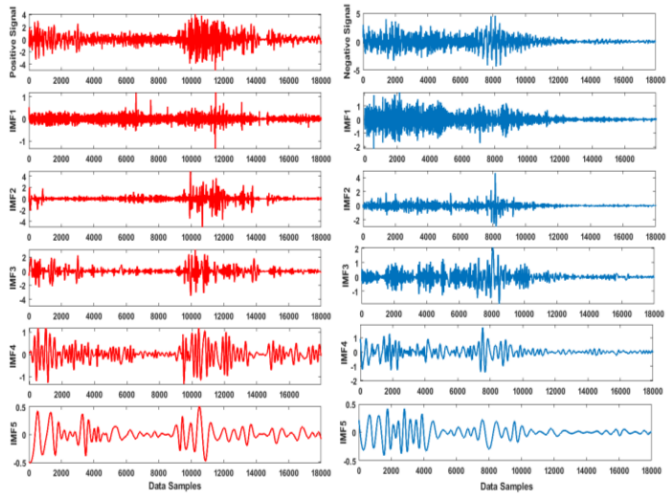

Figure 1: COVID19(+) and COVID19(-) cough acoustic data and 5-layer IMF signals.

## II. METHODS

*A. Dataset*

Cough acoustic signal data for COVID19(+) and COVID19(-) were retrieved from the free accessible website virufy.org. The cough signals were supplied through a mobile app developed via Stanford University. The cough acoustic signals belong to a sum of 1187 people. Whole cough data were specified as negative and positive according to the outcomes achieved from the RT-PCR test. Thereupon, cough signal data of 595 COVID19(+) and 592 COVID19(-) people has been tagged. Whole cough signal data were done away with from the noise. Additionally, all signal data has been preprocessed using z-score normalization technique earlier the research done.

*B. Z-Normalization*

Z-score converts mean of variable to 0 and its standard deviation to 1. To do this, simply subtract the mean and divide by the standard deviation. The z-score normalization could be computed as [9]:

$$z = \frac{x - mean}{std} \qquad (1)$$

Where *x* indicates any point in the dataset, *mean* indicates average of the dataset, and *std* indicates the standard deviation of the dataset.

*C. Empirical Mode Decomposition*

EMD is a suitable technique for analyzing stationary and non-linear series and oscillation signals at the local oscillation level are used. IMFs are decomposed into some oscillation structures in signals [10]. In the EMD algorithm, local peaks are first found in the input signal. By taking a 3rd degree curve at the local maximum point, an overwrap curve is created. By taking a 3rd degree curve at the local minimum point, a sub winding curve is created. The averages of the upper and lower winding curves are calculated. By subtracting these averages from the input signal, the low frequency component of the signal is eliminated. If the signal we get is IMF, the transaction is terminated. If the IMF is not, the processes are repeated for the new signal. To see if the signal is IMF, we look at the IMF conditions. In the first of the IMF conditions, the zero crossings of the signal are equal to or one more than the number of peaks. In the second condition, the winding curves determined by the local minimum and maximum are symmetrical.

*D. Support Vector Machines*

The support vector machine (SVM) is capable of separating data into two or more classes with linear separation mechanisms in two-dimensional space, planar in three-dimensional space, and hyperplane-shaped separation mechanisms in multidimensional space. The basic idea of the SVM is to construct an optimal separator hyperplane on linearly separable data. This idea was put forward by Vapnik in 1995 [11].

*E. Logistic Regression*

Logistic regression (LR) is a technique utilized to detect the reason and result relationship with the explanatory variables in cases where the response variable is observed in categorical, binary, triple and multiple classes. LR is a regression technique where the expected values of the response variable are obtained as probabilities according to the explanatory variables. In simple and multiple regression method, dependent variable should show normal distribution, independent variables should show normal distribution and error variance should show normal distribution with $\varepsilon \cong N(0, \sigma^2)$ parameter. Simple or multiple regression analyzes cannot be applied to data sets that do not contain these conditions. LR is a statistical process that provides the opportunity to classify according to probability rules by computing the estimated values of the dependent variable as probabilities.

*F. Ensemble Bagged Trees*

Ensemble learning is a classification method that uses some learner models together. Bagging is a method that randomly generates data sets and trains models in parallel. It combines models with voting. It is an example algorithm for the random forest (RF) bagging process. In RF, decision trees are randomly generated and trained in parallel. The results are combined with voting to form a forest.

*G. K-Nearest Neighbors*

The k-Nearest Neighbor (kNN) estimation method is a nonparametric classification and regression algorithm. Ease of application and having a simple mathematical basis have made the use of forecasting models widespread in many different fields [12, 13]. It is based on the idea that the outcome of a case is the same as the outcome of its nearest neighbor cases. By means of the training set based on past observations, the dependent variables that are the result of each element (case) of the data set are determined. Forward estimates will be equal to the mean of the results of the current cases of the closest elements in the training dataset. Usually, the closest observations are defined as those with the smallest Euclidean distance to the data point under consideration. The Euclidean distance between the observations can be found depending on the linear distance $x_i$ in the x-plane and the linear distance $y_i$ in the y-plane in the example of the 2-dimensional solution set.

$$\text{Euclidean distance} = \sum_{i=1}^{k}(x_i - y_i)^2 \qquad (2)$$

Model optimization is required to determine $k$, which represents the number of neighbors to consider. As the number of neighbors' k increases, although the calculation step increases, the vulnerability to noise in the training set will decrease and the fit to the test data will increase.

### H. Linear Discriminant Analysis

Discriminant analysis is a multivariable statistical technique that allows an individual or an object to be assigned to one of a finite amount of known different populations according to its measured properties [14]. In discriminant analysis, the discrimination function is obtained according to some assumptions made on the masses. There are different discriminant functions depending on whether the variance-covariance matrices of the populations from which the samples are taken are equal or not. When the variance covariance matrices of the masses are equal, the linear discriminant function is obtained, while if they are different, the square discriminant function is obtained.

### İ. Performance Metrics

In this study, the outcomes were evaluated utilizing five distinct performance measures [15, 16]. These:

$$Accuracy(Acc) = \frac{TP + TN}{TP + FN + FP + TN} \qquad (3)$$

$$Recall(Rec) = \frac{TP}{TP + FN} \qquad (4)$$

$$Specifity(Spe) = \frac{TN}{TN + FP} \qquad (5)$$

$$Precision(Pre) = \frac{TP}{TP + FP} \qquad (6)$$

$$F1 - score(F1) = \frac{2 * PRE * REC}{PRE + REC} \qquad (7)$$

TP shows the people, who are COVID19(+), number and identified as COVID19(+) by the classifier, FN is the number of people who are wrongly stated as COVID19(-), TN the people number, who are actually COVID19(-), and the classifier were stated them as COVID19(-), and FP shows the people number, who are mistakenly identified as COVID19(+) [17].

### III. EXPERIMENTAL RESULTS

In this research work, all the processes have been experimented via MATLAB 2021a software. Z-normalization technique was utilized as preprocessing method. Then, 45 IMF based measures were obtained. The comparison of the classification algorithms is shown in Table 1. As shown in the Table 1, the highest rate of accuracy was obtained as 90.56% using Ensemble Bagged Trees.

Table 1: Performance comparison of classification methods for all EMD based features.

| Algorithm | Performances (%) | | | | |
|---|---|---|---|---|---|
| | Acc | Rec | Spe | Pre | F1 |
| Ensemble Bagged Trees | 90.56 | 90.54 | 90.59 | 90.54 | 90.54 |
| SVM Linear | 89.64 | 88.34 | 90.92 | 90.64 | 89.48 |
| Logistic Regression | 89.39 | 88.18 | 90.59 | 90.31 | 89.23 |
| Linear Discriminant Analysis | 88.88 | 86.32 | 91.43 | 90.93 | 88.56 |
| Medium kNN | 87.95 | 88.68 | 87.23 | 87.35 | 88.01 |

### IV. DISCUSSION

No doubt one of the most researched topics in the last few years is the coronavirus epidemic, which affected all over the world. The most vital point in overcoming this epidemic is the process of making the correct diagnosis. Therefore, there are many machine learning-based researches in the literature.

COVID19(+) detection on cough signal data, which suggests an extraordinary approach, has lately taken its place among the current alternative methods. Performance metrics comparisons of those classification methods among the studies conducted in this topic are given in Table 2. Traditional machine learning approaches and feature extraction processes were generally utilized in these studies. In some studies, in addition to traditional machine learning approaches, deep learning approaches was used to detect COVID19(+) people based on cough sounds. In this study, the performances of five different classification methods were analyzed over the features taken utilizing traditional machine learning approaches. With these five alternative approaches, which work with a very high success rate, a decision support mechanism was recommended to doctors for the detection of COVID19(+) people. The confusion matrices which belongs to classification algorithms are given in Figure 2. Highest accuracy rate belongs to bagged trees classification algorithm shown in Figure 2 (e).

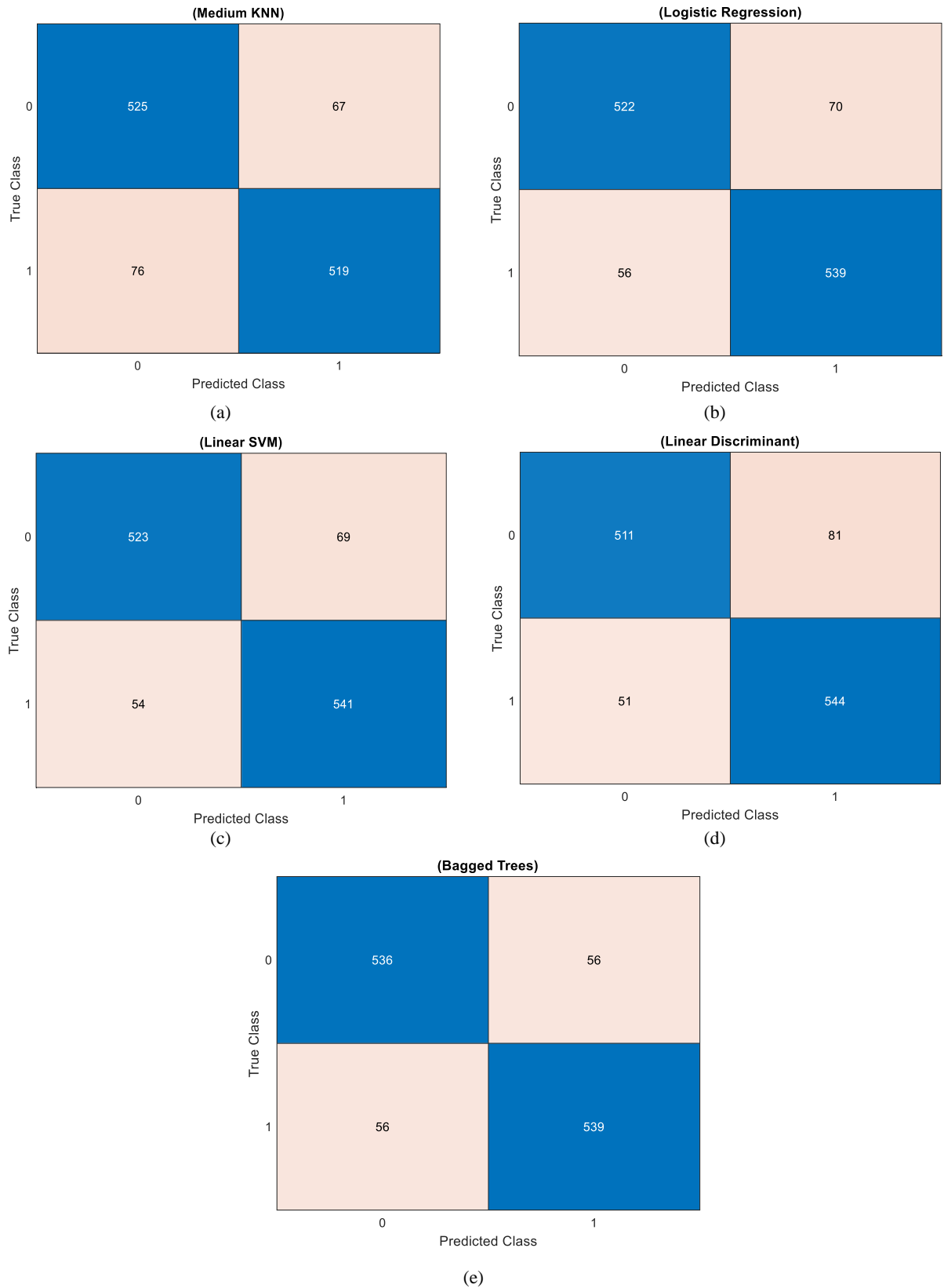

Figure 2: Confusion matrices belonging to the classification methods a) Medium kNN b) Logistic Regression c) Linear SVM d) Linear Discriminant e) Bagged Trees (0, COVID19(−), 1, COVID19(+)).

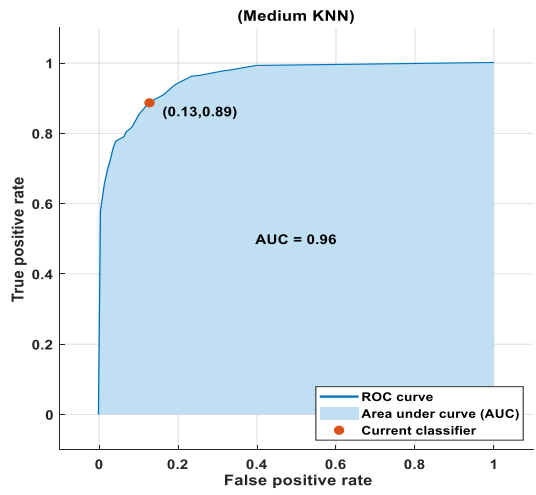
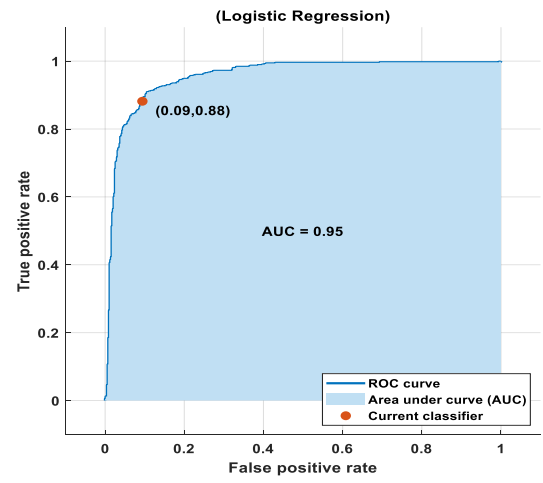
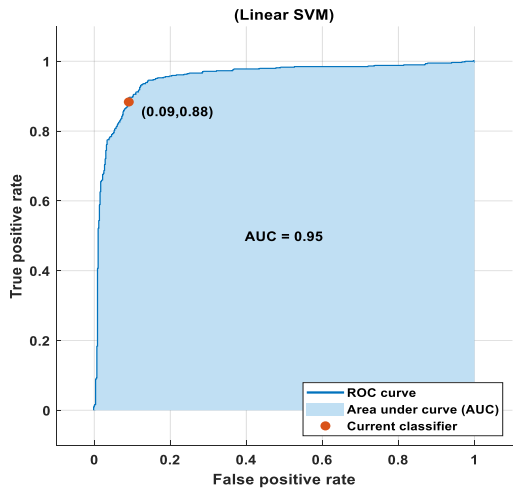
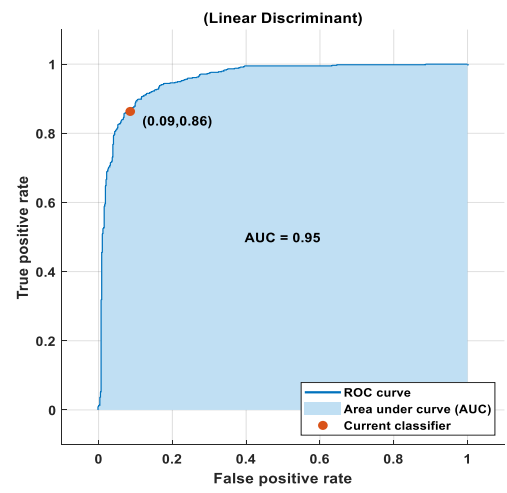
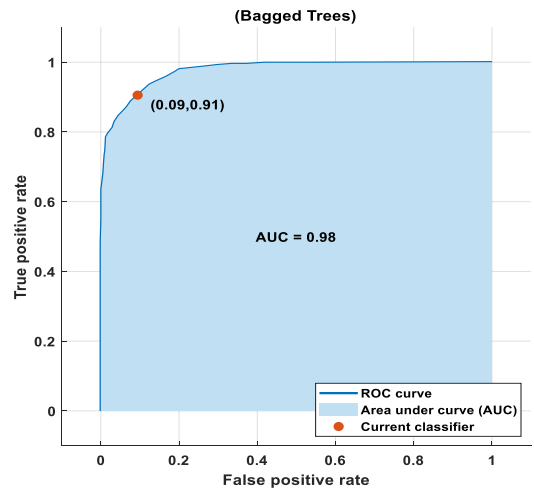

Figure 3: ROC curves belonging to the classification methods a) Medium kNN b) Logistic Regression c) Linear SVM d) Linear Discriminant e) Bagged Trees.

Table 2: COVID19 detection engaging cough acoustic signals in the literature. (Acc, accuracy, Spe, specificity, Rec, recall, F1, F1-score).

| Authors | Methods and Classifiers | Number of Data | Performance (%) |
|---|---|---|---|
| Sharma et al. (2020) [6] | MFCCs, and other spectral measurements/random forest | 941 | Acc = 67.7 |
| Imran et al. (2020) [7] | Mean Score + Mel-frequency cepstral coefficients and Principal component analysis / Support Vector Machines | 543 | Rec=96.0 Spe=95.2 Acc=95.6 F1=95.6 |
| Erdogan and Narin (2021) [8] | Z-Score + Intrinsic Mode Functions and Discrete Wavelet Transform features + Support Vector Machines | 1187 | Rec=99.5 Spe=97.4 Acc=98.4 F1=98.6 |
| Erdogan and Narin (2021) [8] | Z-Score + ResNet50 basis deep features + Support Vector Machines | 1187 | Rec=98.5 Spe=97.3 Acc=97.8 F1=98.0 |
| **This study** | **EMD basis features + Z Score / Ensemble Bagged Trees** | **1187** | **Rec=90.5 Spe=90.6 Acc=90.6 F1=90.5** |

As it can be seen from Figure 2(e), 56 people of those in the COVID19(+) class were detected incorrectly while 539 people were detected correctly. On the other hand 56 people of COVID19(−) class were detected incorrectly while 536 people of those were detected correctly. The receiver operating characteristic curve (ROC) is given in Figure 3. Area Under the ROC Curve (AUC) value is quite high as shown in Figure 3.

In addition to the detection of COVID19(+) with imaging methods, it is of great importance to detect these people with cough-based acoustic sound analysis. With this method, the detection of COVID19(+) can be easily achieved via a smartphone or computer application. With this application, the pandemic can be overcome more easily. From this point of view, it is of great importance that even a single person can be protected from the epidemic during the pandemic. We think that such systems showing high performance will be of importance during pandemic period. One of the most critical restrictions of this research is the restricted number of data. By enhancing the number of data near future, it is thought that the system will be successful on high data numbers. In future studies, it is planned to increase the number of nonlinear measurements.